\documentstyle[prb,psfig,epsfig,aps]{revtex}
\input{epsf}
\begin{document}
\draft
\title{Breathers in Josephson junction ladders:\\
resonances and electromagnetic waves spectroscopy}
\author{A. E. Miroshnichenko, S. Flach and M. V. Fistul}
\address{Max-Planck-Institut f\"ur Physik komplexer Systeme, N\"othnitzer
Strasse 38, D-01187 Dresden, Germany}
\author{Y. Zolotaryuk}
\address{Group of Mathematical Physics, IMM, Technical University of Denmark,
Building 321, Richard Petersens Plads, DK-2800 Kgs. Lyngby, Denmark}
\author{J. B. Page}
\address{Department of Physics and Astronomy, Arizona State University
Tempe, AZ 85287-1504, Arizona, USA}
\date{\today}
\wideabs{
\maketitle
\begin{abstract}
We present a theoretical study of the resonant interaction between 
dynamical localized states ({\it discrete breathers}) 
and linear electromagnetic excitations (EEs) 
in Josephson junction ladders. 
By making use of direct 
numerical simulations we find that such an interaction manifests itself by
{\it resonant steps} and various sharp switchings ({\it voltage jumps}) in the 
current-voltage characteristics. Moreover, the power of ac oscillations 
away from the breather center (the {\it breather tail}) displays 
singularities as the externally applied dc bias decreases. All these features 
can be mapped to the spectrum of EEs that has been derived analytically
and numerically.  
Using
an improved analysis of the breather tail, a spectroscopy of the 
EEs is developed. The nature of breather instability
driven by localized EEs is established. 
\end{abstract}
}

\section{Introduction}
Various nonlinear and discrete systems (nonlinear lattices) have attracted a 
lot of interest as they display diverse fascinating phenomena. \cite{shs94}
Well known examples of such phenomena are solitary excitations, 
propagation of (non)linear waves and appearance of various 
inhomogeneous structures.

Moreover, the interest in this area was boosted by the prediction, 
theoretical analysis \cite{ajsjbp95,sa97,sfcrw98} and the 
subsequent observation
\cite{bisjabsplgfsapsarbwzwmis99,utslqeajs99,etjjmtpo00,pbdaavusfyz00,pbdaavu00,etjjmabtpo00}  
of {\it intrinsic dynamic localized excitations} (discrete breathers) 
that are periodic in time and localized in space. 
Note here, that the origin of such dynamical localization is not the 
presence of disorder but the interplay between the nonlinearity and 
discreteness.

These peculiar states have been experimentally verified
as vibrational modes in low-dimensional crystals,
\cite{bisjabsplgfsapsarbwzwmis99}  
localized excitations in spin lattices, \cite{utslqeajs99}  and 
localized resistive states in Josephson junction arrays 
\cite{etjjmtpo00,pbdaavusfyz00,pbdaavu00,etjjmabtpo00}.
The latter systems are of special interest because they have served for many years 
as well-controlled laboratory objects to study various nonlinear 
phenomena. \cite{shs94,kkl86} Moreover, at variance 
with other systems  
intrinsic localized modes found in Josephson coupled systems can be 
excited 
in the presence of time independent external driving forces.

\begin{figure}[htb]
\vspace{10pt}
\centerline{\psfig{figure=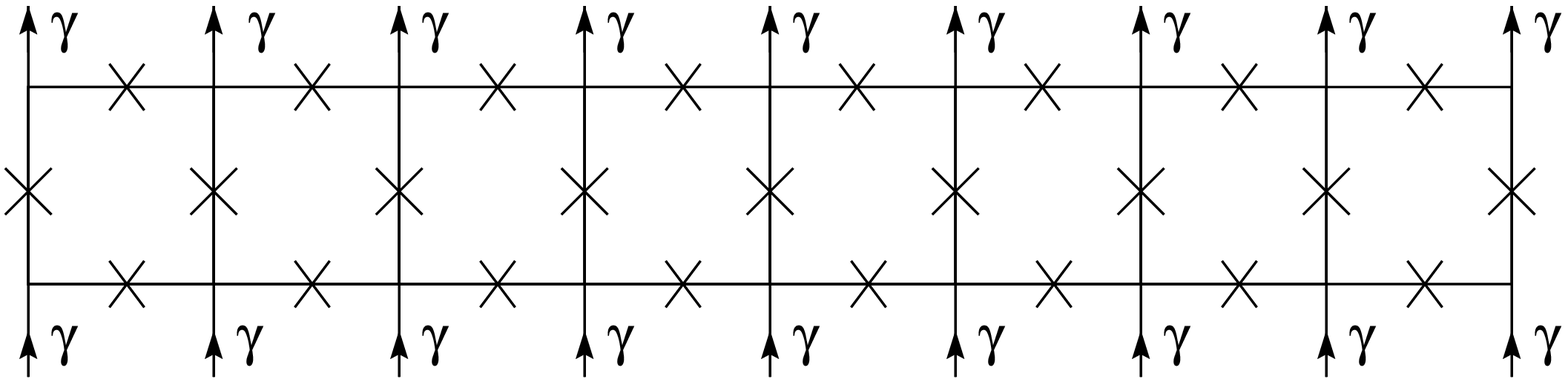,width=82mm,height=20mm}}
\vspace{2pt}
\caption{
Josephson junction ladder. Crosses mark the individual junctions.
Arrows indicate the direction of external current flow (dc bias $\gamma$).
}
\label{fig1}
\end{figure}

A well known structure where dynamical 
localized states appear, 
is the anisotropic Josephson junction ladder (JJL) 
\cite{etjjmtpo00,pbdaavusfyz00,pbdaavu00,etjjmabtpo00}.  
A schematic view of such a ladder is given
in Fig.\ref{fig1}. 
The ladder contains small Josephson junctions indicated by  
crosses in Fig.\ref{fig1}, 
in both  longitudinal ({\it vertical} junctions) and 
transverse ({\it horizontal} junctions) directions to the dc bias current 
$\gamma$. 
The anisotropy 
of the ladder is due to the different sizes of vertical and 
horizontal junctions and is characterized by the anisotropy parameter 
$\eta~=~\frac{I_{cH}}{I_{cV}}$, where $I_{cV}$ and $I_{cH}$ are respectively 
the critical currents of the vertical and horizontal junctions.
%

Dynamical localized excitations persist 
in a JJL
due to the intrinsic 
bistability property of a single small underdamped Josephson junction. 
One of these stable states is a superconducting one with zero voltage drop
across the junction. The other state is a resistive one with a nonzero
voltage drop, also called whirling state. 
A breather state in a JJL is characterized
by a few junctions being in the resistive state, while the rest of all 
junctions are in the superconducting state. 
The presence of breather states 
can be verified by measurements of a total dc voltage drop across the ladder,
which is used to plot current-voltage ($I-V$) characteristics.
This method does not provide spatially resolved information. It has been successfully
combined with snapshots made using low temperature laser microscopy techniques, 
\cite{pbdaavusfyz00}
which allow for a spatial resolution of the dc voltage drops. 
Note that both methods provide only with time-averaged voltage drop
data, so the internal
dynamics are so far not accessible in experiments.

However, the full dynamical picture is much
more subtle than the time-averaged picture might suggest. In particular
the Josephson junctions in the superconducting state exhibit
small librations of 
the Josephson phase and correspondingly, nonzero ac voltage drops.
The amplitude of these librations should decay to zero with increasing
distance from the resistive junctions. 
It is also well known that various systems 
of coupled Josephson junctions support 
a delocalized class of excitations, namely, 
small amplitude electromagnetic waves 
(EWs). \cite{kkl86,pcmvfavubamsf99} In JJLs 
the spectrum of EWs consists of three branches and 
depends in a complex manner on the
anisotropy $\eta$ and the dc bias $\gamma$. 
The resonant interaction of these EWs with the homogeneous 
whirling state (HWS) in the presence of an externally applied magnetic field 
has been studied in [Ref.~\onlinecite{pcmvfavubamsf99}]. It was shown that such an 
interaction leads to {\it resonant steps} in $I-V$ curves and 
the voltage positions of resonant steps can be mapped onto the spectrum of EWs.

Early theoretical studies \cite{sfms99} dealt with the possibility of 
resonant interaction of breather states with EWs of the ladder.
Due to the intrinsic spatial inhomogeneity of a breather state, 
EWs can be excited even in the absence of an externally applied magnetic field. 
The resonant interaction of EWs with the 
breather state manifests itself through the appearance of
resonant steps \cite{etjjmabtpo00} 
and various switchings between 
different breather states ({\it "voltage jumps"}) in $I-V$ curves.
Moreover, the amplitude of the Josephson phase librations 
at some distance from the breather center increases drastically 
due to this interaction. \cite{sfms99} 

The spatial inhomogeneity of a breather state allows also
for the appearance of {\it localized} small amplitude 
electromagnetic excitations (EEs). 
We will show that the
presence of localized EEs and their resonant interaction with the breather
is of crucial importance.
Our study resolves a long standing puzzle of the nature of
breather instability. We show that most of these instabilities
are driven by localized EEs. Especially the so-called retrapping
(i.e. the switching from a breather to the superconducting
state) is due to these instabilities and can not be explained
using standard retrapping arguments.

In this paper we present a consistent theoretical study of resonant 
interactions between the breather states and EEs. 
We will derive the spectrum of EWs and calculate the ac power of  
oscillations at some distance from the breather center ("breather tail"). 
By making use of direct
numerical simulations of the dynamics of JJLs in different parameters
ranges, 
we demonstrate that the $I-V$ curves
display a variety of different
resonance steps and voltage jumps as the external dc bias, and correspondingly 
the breather frequency, decrease. All these features are mapped onto 
various resonances of the breather
frequency (or even its second and third harmonics) 
with EE frequencies, as well as with combinations
of frequencies of different EEs. 
Moreover, by monitoring the power of ac oscillations 
as a function of the dc bias
 $\gamma$, we develop a spectroscopy of the EWs in the JJL.

The paper is organized as follows:
In section II we derive the equations of motion within the framework of the 
Resistively Shunted Junction (RSJ) model 
\cite{kkl86} and obtain
the spectrum of linear EWs of a JJL. 
A symmetry classification of different types of breathers is presented
in section III. An improved analysis of the breather tail and the corresponding 
dependence of the
power of ac oscillations on the breather frequency 
is given in section IV. In section V we will classify different
resonances of breathers with EEs. 
Direct numerical simulations of the dynamics of breather states 
are presented and discussed in section VI. Section VII is devoted
to the interpretation of the obtained resonances and switchings.

\section{Dynamics of a JJL 
and the spectrum of linear electromagnetic waves}

The complete dynamics of a JJL is 
determined by the time dependent Josephson phases 
of vertical $\phi_n^v$, upper horizontal 
$\phi_n^h$ and lower horizontal $\tilde{\phi}_n^h$ junctions. 
The subscript $n$ labels the cell number.
By making use of the RSJ model for each junction 
\cite{kkl86} we obtain the following set of equations:
\begin{eqnarray}\label{2-1}
{\cal N}(\phi_n^v)=I_n^v\nonumber \\
{\cal N}(\phi_n^h)=\frac{1}{\eta}I_n^h\\
{\cal N}(\tilde{\phi}_n^h)=\frac{1}{\eta}\tilde{I}_n^h\nonumber\;,
\end{eqnarray}
where the nonlinear operator ${\cal N}$ is defined as
\begin{eqnarray}
{\cal N}(y)=\ddot{y}+\alpha\dot{y}+\sin y\;.
\end{eqnarray}
Here, the unit of time is the inverse plasma frequency $\omega_p^{-1}$, and 
the currents $I_n^v$, $I_n^h$ and $\tilde{I}_n^h$ are measured in
units of the 
critical current of vertical junctions. 
The dimensionless parameter $\alpha$ determines the damping strength 
in each junction.
Note that the positive current direction is chosen
to be directed from bottom to top and 
from left to right.
The currents flowing via the Josephson junctions and the Josephson phases 
are governed by 
the Kirchhoff laws 
\begin{eqnarray}
\gamma=I_n^v+I_n^h-I_{n-1}^h\nonumber\\
\gamma=I_n^v-\tilde{I}_n^h+\tilde{I}_{n-1}^h\;,
\end{eqnarray}
and the flux quantization law in each cell:
\begin{eqnarray}\label{meshc}
-\beta_L I_n^m=\phi_n^h+\phi_{n+1}^v-\tilde{\phi}_n^h-\phi_n^v\;.
\end{eqnarray}
Here, we introduced the mesh currents $I_n^m$ and the normalized 
inductance of the cell, $\beta_L$. 
 
The Kirchhoff equations can be subtracted from each other
yielding
\begin{eqnarray}
I_n^h+\tilde{I}_n^h=C\;,
\end{eqnarray}
where $C$ is a constant for the whole ladder. 
This constant corresponds to the net difference between the currents 
flowing through the upper and lower horizontal junctions.
For an open  ladder of finite size, $C$ is zero. For
a ladder of annular geometry with periodic boundary conditions 
$C$ may be 
nonzero
and corresponds to the flux "trapped" by the ladder ring. 
In the following we
will consider the case of a finite open ladder with $C=0$. 
Then we may always eliminate the 
currents $\tilde{I}_n^h$ from the
set of equations (\ref{2-1}) as
$\tilde{I}_n^h=-I_n^h$.
Since the junction width is larger than the London penetration depth, the
mesh currents are \cite{jrphsjzjwtpo93}
\begin{eqnarray}\label{2-6}
I_n^m=I_n^h\;,
\end{eqnarray}
and the currents flowing through the vertical junctions are expressed as
\begin{eqnarray}\label{2-7}
I_n^v = \gamma-I_n^m+I^m_{n-1}\;.
\end{eqnarray}

Inserting the relations (\ref{meshc}), (\ref{2-6}), and (\ref{2-7})
into (\ref{2-1}) we finally obtain
the following set of coupled differential equations 
\cite{pbdaavusfyz00,pcmvfavubamsf99,sfms99,gggfspug96} :
\begin{eqnarray}\label{2-8}
   \ddot{\phi}_{n}^v+\alpha\dot{\phi}^v_{n}+\sin\phi_{n}^v&
   =&\gamma+\frac{1}{\beta_{L}}(\triangle\phi_{n}^v+\nabla\phi_{n-1}^h
   -\nabla\tilde{\phi}_{n-1}^h)\nonumber \\
   \ddot{\phi}_{n}^h+\alpha\dot{\phi}_{n}^h+\sin\phi_{n}^h &
   =&-\frac{1}{\eta\beta_{L}}(\nabla\phi_{n}^v+\phi_{n}^h-\tilde{\phi}_{n}^h) 
   \\
  \ddot{\tilde{\phi}}{}^h_n+\alpha\dot{\tilde{\phi}}{}^h_n+
  \sin\tilde{\phi}_n^h&=&\frac{1}{\eta\beta_{L}}
  (\nabla\phi_{n}^v+\phi^h_n-\tilde{\phi}_{n}^h)\nonumber\;, 
\end{eqnarray}
where we use the notations $\triangle f_n\equiv f_{n-1}-2f_n+f_{n+1}$ 
and $\nabla f_n\equiv f_{n+1}-f_n$.

Next, we carry out the analysis of the {\it delocalized class} of excitations, 
namely of small amplitude electromagnetic waves (EWs). Note here that the 
spectrum of EWs $\omega(q)$ depends crucially on the state of the
system. In the following we will consider a static (superconducting) 
state, $\phi_n^{*v}=\arcsin\gamma$ and
$\phi_n^{*h}=\tilde{\phi}_n^{*h}=0$.   
We decompose the Josephson phases into the particular form:
\begin{eqnarray}
\phi_n^v=\phi^{*v}_n+\varphi_n^v\nonumber\\
\phi_n^h=\phi_n^{*h}+\varphi_n^h\\
\tilde{\phi}_n^h=\tilde{\phi}_n^{*h}+\tilde{\varphi}_n^h\nonumber\;,
\end{eqnarray} 
where $\varphi_n^v$, $\varphi_n^h$ and $\tilde{\varphi}_n^h$ describe the 
small amplitude EWs. 
Substituting these expressions into system (\ref{2-8}) and using the
smallness of the amplitude of EWs
we obtain
\begin{eqnarray}\label{2-10}
\ddot{\varphi}_{n}^v+\alpha\dot{\varphi}^v_{n}+\sqrt{1-\gamma^2}\varphi_{n}^v&=&\frac{1}{\beta_{L}}(\triangle\varphi_{n}^v+\nabla\varphi_{n-1}^h-\nabla\tilde{\varphi}_{n-1}^h)\nonumber \\
   \ddot{\varphi}_{n}^h+\alpha\dot{\varphi}_{n}^h+\varphi_{n}^h  
&=&-\frac{1}{\eta\beta_{L}}(\nabla\varphi_{n}^v+\varphi_{n}^h-\tilde{\varphi}_{n}^h) \\
\ddot{\tilde{\varphi}}{}_{n}^h+\alpha\dot{\tilde{\varphi}}{}_{n}^h+\tilde{\varphi}^h_n&=&\frac{1}{\eta\beta_{L}}(\nabla\varphi_{n}^v+\varphi_{n}^h-\tilde{\varphi}_{n}^h)\nonumber\;. 
\end{eqnarray}
In the weakly damped case as the parameter $\alpha~\ll~ 1$ we can derive the
spectrum $\omega(q)$ of EWs neglecting effects of dissipation.
By taking the Josephson phases 
$\varphi_n^v$, $\varphi_n^h$ and $\tilde{\varphi}_n^h$ in the form
\begin{eqnarray}
\left(\begin{array}{c}\varphi_n^v \\ \varphi_n^h \\
\tilde{\varphi}_n^h\end{array}\right)=e^{i(qn+\omega
t)}\left(\begin{array}{c}\Delta_v \\ \Delta_h \\
\tilde{\Delta}_h\end{array}\right)\;,
\end{eqnarray}
we find that the spectrum consists of three branches.
The first is given by 
\begin{eqnarray}
\omega_0^2=1,\;\; & \Delta_v=0, \;\; & \Delta_h=\tilde{\Delta}_h\;.
\end{eqnarray}
This branch is dispersionless and
EWs corresponding to this branch are characterized by
nonactive vertical junctions and in phase (symmetric) librations of 
the Josephson phases of upper and lower horizontal junctions.

The two other solutions are generalizations of 
those discussed in [Ref.~\onlinecite{sfms99}], namely
\begin{eqnarray}\label{2-13}
\omega_{\pm}^2&=&F\pm\sqrt{F^2-G}\;,\nonumber\\
F&=&\frac{1}{2}+\frac{1}{\beta_L\eta}+\frac{1}{2}
\sqrt{1-\gamma^2}+\frac{1}{\beta_L}(1-\cos q)\;,\\
G&=&(1+\frac{2}{\beta_L\eta})\sqrt{1-\gamma^2}+\frac{2}{\beta_L}(1-\cos
q)\nonumber\;.
\end{eqnarray}

Both branches have a nonzero dispersion. 

The branch $\omega_{+}$ is characterized by $\Delta_h=-\tilde{\Delta}_h$ for all wave vectors 
$q$, i.e. the upper and lower
horizontal phases are antisymmetric. The frequency range
of the branch is above the degenerate branch $\omega_0$,
i.e. $\omega_{+}(q)>\omega_0$ and it depends strongly on
$\beta_L$. As the parameter  $\beta_L$ increases, the
width of $\omega_{+}(q)$ decreases
and the branch approaches the 
dispersionless one, $\omega_0$. In the opposite case of
small $\beta_L$, the frequencies $\omega_{+}(q)$ increase as
$1/\sqrt{\beta_L}$. 
For zero wave number $q=0$, the amplitudes of EWs in this branch
are characterized by
$\Delta_v=0$ and $\Delta_h=-\tilde{\Delta}_h$, which means that only horizontal junctions are
excited.

The branch $\omega_{-}$ becomes dispersionless for the particular case 
of $\gamma=0$ 
and it corresponds
to the dispersionless band obtained in [Ref.~\onlinecite{sfms99}].
The frequency range of this branch is located
below $\omega_0$ i.e. $\omega_{-}(q)<\omega_0(q)$. 
For zero wave number $q=0$ the horizontal junctions are not active 
($\Delta_h=\tilde{\Delta}_h=0$) and only vertical junctions are excited. 

For a finite size ladder with open boundary conditions and
$N$ cells, i.e. $N+1$ vertical junctions, the spectrum of linear
waves is discrete and characterized by the following choice of
allowed wave number values:
\begin{equation}
q_l=\frac{l\pi}{N+1}\;\;,\;\;l=0,1,2,...,N\;.
\label{obc}
\end{equation}
These EWs are {\it cavity modes} of the JJL. Odd values of $l$ 
correspond to antisymmetric eigenvectors (with respect
to reflections at the center of the ladder), whereas
even values correspond to symmetric ones.  

Note that the above spectrum 
of EWs (\ref{2-13})
is in general quite different from the EW spectrum of the homogeneous whirling  
state. \cite{pcmvfavubamsf99}
The latter can be obtained by choosing $\gamma=1$ in 
(\ref{2-13}). The main difference is the appearance of a
gapless (acoustic) lower branch.

\section{Symmetries and dc bias dependent frequencies of breathers}

In this Section we turn to the analysis of {\it dynamic localized excitations}
(breathers) in JJLs. As mentioned in the introduction, breather states
correspond to a few junctions being in the resistive state with all other
junctions being  
in the superconducting state. The Josephson phases of 
resistive junctions are unbounded in time and the Josephson phases 
of superconducting junctions display small amplitude librations with 
a frequency $\Omega$. This frequency is called the {\it breather frequency}.

Experiments
\cite{etjjmtpo00,pbdaavusfyz00,pbdaavu00,etjjmabtpo00}
have revealed many different breather structures.
All of them can be
classified into three symmetry types using the reflection 
symmetries of the JJL. Some possible realizations
are presented schematically in Fig. 2.

%
\begin{figure}[htb]
\vspace{20pt}
\centerline{\psfig{figure=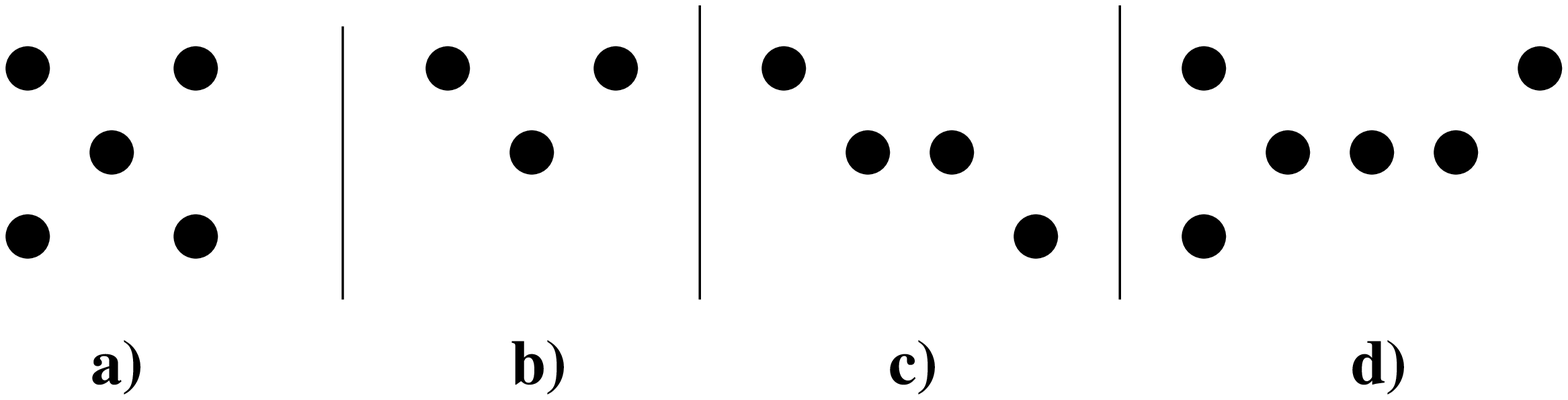,height=16mm,width=82mm}}
\vspace{2pt}
\caption{
Examples of different types of breathers: 
a) up-down symmetry, b) left-right symmetry, c) inversion
symmetry, d) no symmetry. Black spots indicate the positions of 
whirling junctions.}
\label{fig2}
\end{figure}

Breathers from the first group reveal an {\bf 'up-down' reflection symmetry} 
$\hat{S}_{ud}$
(see, Fig.\ref{fig2}a),
i.e. they are invariant under exchange of upper and lower horizontal
junctions. The second group consists of breathers invariant
under a {\bf 'left-right' reflection symmetry} $\hat{S}_{lr}$ (see, Fig.\ref{fig2}b), 
i.e. they are invariant
under a reflection at a vertical line cutting the ladder (this line
is located either in the middle between two vertical junctions, or
passes directly through one vertical junction). 
A third distinct group of breathers possesses an {\bf 'inversion' symmetry}
$\hat{S}_{in}$ (Fig.\ref{fig2}c), 
i.e. these breathers
are invariant under a reflection at a point which is either located
on a vertical junction or in the center of a plaquette. A fourth group 
of breathers has {\bf no symmetries} at all and does not belong to any
of the three listed symmetry types. A particular example of a breather 
without symmetry is 
shown in Fig.\ref{fig2}d. 
All of these types of breather excitations  have been
observed experimentally
and numerically. 
\cite{etjjmtpo00,pbdaavusfyz00,pbdaavu00,etjjmabtpo00} 
Each group of breathers can also have a different number of vertical junctions 
in the resistive state. Note that 
the particular example in Fig.\ref{fig2}a
possesses not only $\hat{S}_{ud}$ symmetry, but also $\hat{S}_{lr}$
and $\hat{S}_{in}$ symmetries. However it is also possible to construct
more complex breather states which display $\hat{S}_{ud}$ symmetry only.

Next we derive the average voltage drop across a resistive
vertical junction $V=\langle \dot{\phi}^v \rangle$ for different breather types.
For the particular case of a breather with up-down symmetry $\hat{S}_{ud}$,
there are $k$ rotating vertical junctions in cells $(i+1)..(i+k)$ and 
two rotating horizontal junctions in the $(i)$th and $(i+k)$th cells
respectively.
By making use of the flux quantization law (\ref{meshc}) we obtain identical
voltage drops across the vertical junctions in the resistive state.
For the same reasons the voltage drops
across the horizontal junctions are two times smaller. 
Neglecting nonlinear contributions from the time average 
of $\sin \phi$ on resistive junctions 
we obtain:
\begin{equation}
 \begin{array}{lcr}
 \alpha\frac{V}{2}=\frac{1}{\eta}\langle I^m_{i} \rangle \\
\alpha V=\gamma+\langle I^m_{i+1} \rangle - \langle I^m_{i} \rangle \\
...\\
\alpha V=\gamma+\langle I^m_{i+k} \rangle -\langle I^m_{i+k-1} \rangle \\
\alpha\frac{V}{2}=-\frac{1}{\eta}\langle I^m_{i+k} \rangle\;.
 \end{array}
\end{equation}
Thus, the voltage drop across a resistive vertical junction
that corresponds to the experimentally measured
voltage drop across the ladder, is given by
\begin{equation}
V=\frac{k\gamma}{\alpha(k+\eta)}\;.
\end{equation}
Similarly we analyze a breather with left-right symmetry
$\hat{S}_{lr}$ and with inversion symmetry $\hat{S}_{in}$. 
Taking into account that in these cases the voltage
drops across resistive horizontal and vertical junctions are identical,
we find
\begin{equation}\label{vlr}
V=\frac{k\gamma}{\alpha(k+2\eta)}\;.
\end{equation}
In a similar manner the result for a breather which has no symmetry (cf. 
Fig. 2d) reads
\begin{equation}
V=\frac{k\gamma}{\alpha(k+\frac{3}{2}\eta)}\;.
\end{equation}
The above results for the dependence of 
the average voltage drop on the dc bias
may be combined in a single expression
\begin{equation}\label{eq3}
V=\frac{k\gamma}{\alpha[k+(3-\frac{1}{2}\delta)\eta]}\;,
\end{equation}
where $k$ is the number of vertical rotating junctions and $\delta$ denotes
the number of resistive horizontal junctions. Note that 
$\delta=4$ for breathers with up-down symmetry, 
$\delta=2$ for left-right
or inversion symmetry, and $\delta=3$ for no symmetry.

In order to analyze the interaction between the breather state and the 
linear EEs we need to know the frequency $\Omega$ of a breather solution.
Noting that the breather frequency is given by the lowest realized
voltage drop across a resistive junction, we find 

{\it 1) up-down symmetry}
\begin{equation}\label{3-6}
\Omega=\frac{k\gamma}{2\alpha(k+\eta)}\;,
\end{equation}

{\it 2) left-right symmetry and inversion symmetry}
\begin{equation}\label{3-7}
\Omega=\frac{k\gamma}{\alpha(k+2\eta)}\;,
\end{equation}

{\it 3) no symmetry}
\begin{equation}\label{3-8}
\Omega=\frac{k\gamma}{\alpha(2k+3\eta)}\;.
\end{equation}

\section{Spatial tails of breathers}

We consider the spatial dependence of Josephson phases in the presence of 
a breather state. At some distance from the breather center ("breather tail") 
the Josephson phases librate with small amplitudes. In order to analyze 
the breather dynamics in the tail, we use the linearized system of equations
(\ref{2-10}). 
Keeping in mind the time periodicity of the breather solution, the
librating Josephson phases take the form (for $n~<~0$, and with 
the breather center as the origin) 
\begin{eqnarray}\label{eq9}
\left(\begin{array}{c}\varphi_n^v \\ \varphi_n^h \\
\tilde{\varphi}_n^h\end{array}\right)=e^{\lambda n+i\Omega t}\left(\begin{array}{c}
\Delta_v \\ \Delta_h \\
\tilde{\Delta}_h\end{array}\right)\;.
\end{eqnarray}
Introducing new sum and difference variables of the 
amplitudes of Josephson phases of horizontal junctions   
\begin{equation}
\Delta_h^-=\frac{1}{2}(\Delta_h-\tilde{\Delta}_h)\;\;,\;\; 
\Delta_h^+=\frac{1}{2}(\Delta_h+\tilde{\Delta}_h)\;,
\end{equation}
it follows that
\begin{eqnarray}\label{4-4}
(A-\frac{2}{\beta_L}\cosh\lambda)\Delta_v-\frac{2}{\beta_L}(1-e^{-\lambda})
\Delta_h^-=0\nonumber\\
-\frac{1}{\beta_L\eta}(1-e^{\lambda})\Delta_v+B\Delta_h^-=0\\
(-\Omega^2+i\alpha\Omega+1)\Delta_h^+=0\nonumber\;,
\end{eqnarray}
where the frequency dependent parameters $A$ and $B$ are
\begin{eqnarray}
A&=&-\Omega^2+i\alpha\Omega+\sqrt{1-\gamma^2}+\frac{2}{\beta_L}\nonumber\\
B&=&-\Omega^2+i\alpha\Omega+1+\frac{2}{\beta_L\eta}\;.
\end{eqnarray}
~From (\ref{4-4}) we immediately find that 
$\Delta_h^+=0$ and hence $\Delta_h^-=\Delta_h$. It follows that  
$\varphi_n^h=-\tilde{\varphi}_n^h$, i.e. 
breather tails appear with perfect up-down symmetry. This is at variance with
the complex symmetry properties of the resistive breather center (see the 
Section III).
A nontrivial solution to the first two equations in (\ref{4-4})
exists if
\begin{equation}
AB-\frac{2}{\beta_L}B\cosh\lambda-\frac{4}{\beta_L^2\eta}+
\frac{4}{\beta_L^2\eta}\cosh\lambda=0\;.
\end{equation}
The dependence of the complex parameter 
$\lambda$ on the breather frequency $\Omega$ 
is given by
\begin{eqnarray}\label{4-7}
\lambda=\ln(z+\sqrt{z^2-1})~~,
\end{eqnarray}
with
\begin{eqnarray}\label{4-8}
z~=~\frac{4-\beta_L^2\eta AB}{4-2\beta_L\eta B}\;.
\end{eqnarray}
Note here that this expression can be obtained directly from (\ref{2-13}) 
by assuming that $q=i\lambda$  and substituting $\Omega^2-i\alpha\Omega$ 
instead of $\omega^2_{\pm}$.

\begin{figure}[htb]
\vspace{20pt}
\centerline{\psfig{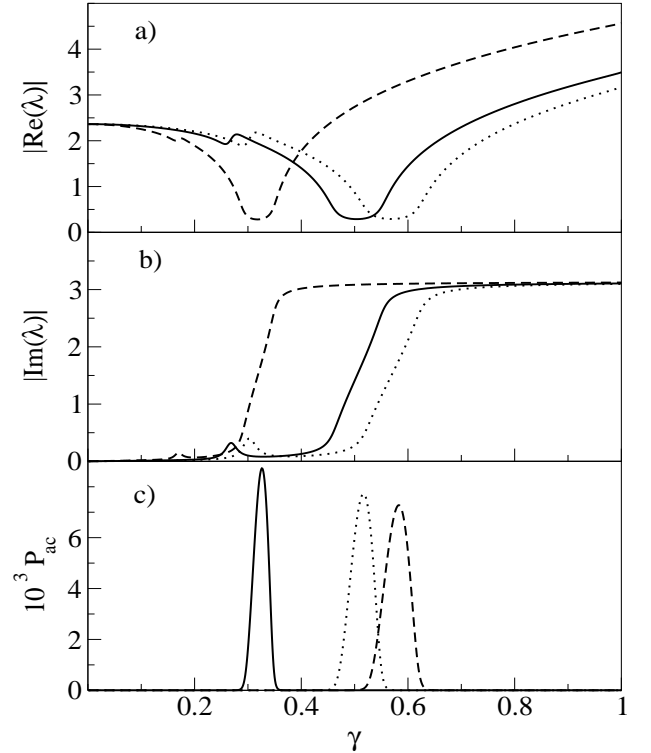}}
\vspace{2pt}
\caption{
Dependence of (a) real part $\Re(\lambda)$, (b) imaginary part $\Im(\lambda)$,
and (c) $P_{ac}$
on $\gamma$ for different types of breathers: solid line - up-down
symmetry, dashed line - left-right symmetry, dotted line - no symmetry. 
The parameters are $\alpha=0.1$, $\beta_L=3.0$, $\eta=0.35$ and $k=1$. 
}
\label{fig3}
\end{figure}

The real $\Re (\lambda)$ and imaginary $\Im (\lambda)$ parts of $\lambda$ 
determine respectively the 
spatial decay and spatial period of oscillations of Josephson phases in the 
breather tail.
Moreover,  $\Re (\lambda)$ and $\Im (\lambda)$ strongly depend on the breather 
frequency that in turn, can be changed by varying the external dc bias 
$\gamma$. In
Fig.\ref{fig3}(a,b) 
we plot the real and imaginary parts of $\lambda$ for three breather types 
(cf. Fig. \ref{fig2})
versus $\gamma$. The minima of the real part of
$\lambda$ correspond to resonances with linear EWs $\omega_{+}(q)$ and
$\omega_{-}(q)$. As one can see, the current positions of these
minima shift for different types of breathers.
Note that the $\gamma$ range in the plots extends from zero to one. In fact 
the breather will exist only in a narrower current region. 
This is due to
the presence of both finite nonzero retrapping current 
\cite{kkl86,abgp82} (switching to 
the
superconducting state) and a particular current $\gamma < 1$, 
where the breather switches to the HWS.  
%

\begin{figure}[htb]
\vspace{20pt}
\centerline{\psfig{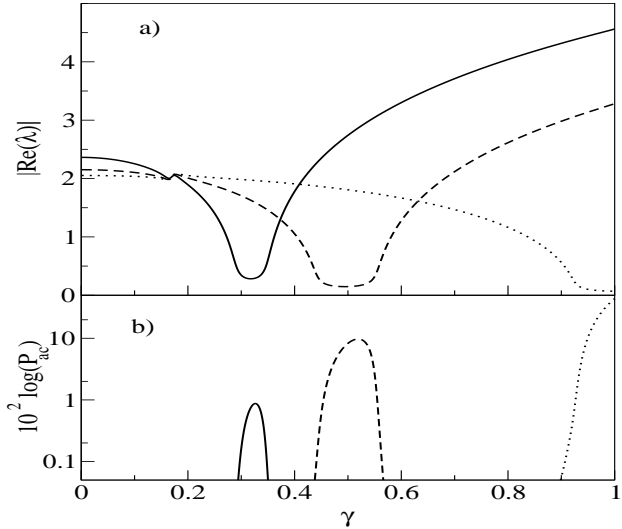}}
\vspace{2pt}
\caption{(a) $\Re(\lambda)$  and (b) $P_{ac}$  versus $\gamma$ for different 
values of the inductance of the cell: 
$\beta_L=3$, solid line; $\beta_L=1$, dashed line;
$\beta_L=0.2$, dotted line. All of these curves are for
a left-right symmetry breather with $k=1$.}
\label{fig4}
\end{figure}

Since the EW frequencies 
$\omega_{+}(q)$ decrease with increasing inductance of the cell 
(\ref{2-13}), 
the position of the global
minimum  also depends strongly on 
$\beta_L$ (see Fig.\ref{fig4}(a)).
%

Although the dynamics of the Josephson phases in the breather tail 
are completely determined by the 
dependence of the parameter $\lambda$ on the breather frequency, 
it may be more convenient
to measure just the time-average power $P_{ac}$ of the libration of a junction 
at the edge of a JJL.
We show below that the monitoring of $P_{ac}$ upon decreasing
the dc bias 
allows one to develop a spectroscopy of EWs.
The value of $P_{ac}$ is determined by the average 
kinetic energy 
of the edge vertical junction 
\begin{eqnarray}\label{pac}
P_{ac}=\frac{1}{2}\big\langle
\dot{\phi}^{v^2}_l\big\rangle\;\;,\;\;l=-\frac{N}{2}
\end{eqnarray}
and is derived
in Appendix A. The typical dependences of  $P_{ac}$ 
on the dc bias and the 
inductance of the cell for different types of breathers are presented in 
Figs. \ref{fig3}c, \ref{fig4}b.

\section{Breather instabilities and classification of resonances}

In early experiments on breathers in JJLs 
\cite{etjjmtpo00,pbdaavusfyz00}
it was observed that 
breather states may switch to other breather states or HWSs
upon variation of the dc bias $\gamma$. 
These switchings can be either
due to the disappearance of the breather state as a solution of the
underlying dynamical equations or due to the novel effect of 
{\it dynamical instability} of the breather state. In the 
latter case the solution continues to exist, but turns unstable,
forcing the system to search for another stable attractor state.
The cause of such instabilities is the resonant interaction of the 
breather state 
with small amplitude (localized or delocalized) EEs.

In  order to analyze the breather instabilities in more precise terms, 
we have to linearize the phase space flow around a given
breather solution and study the corresponding Floquet eigenvalue
problem (see Appendix B). The outcome of this analysis
is a spectrum of Floquet multipliers (i.e. eigenvalues) $\mu$.
In most cases the multipliers reside on a circle
of radius 
\begin{eqnarray}\label{radius}
R(\alpha)=e^\frac{-\alpha \pi}{\Omega}\;,
\end{eqnarray}
in the complex plane, which is less than 1.
Thus, the multipliers can be expressed in the
general form
\begin{eqnarray}\label{floq}
\mu=R(\alpha)\exp\left(\pm i\omega\frac{2\pi}{\Omega}\right)\;,
\end{eqnarray}
where $\Omega$ is the breather frequency and $\omega$ is some
characteristic EE frequency. The corresponding eigenvectors 
may be divided into two classes, namely those which are localized
on the breather and those which are delocalized. While it is
more involved to make analytical predictions for the class
of localized Floquet eigenstates, we may immediately proceed with
the characterization of the class of delocalized eigenstates. 
Although the concrete
form of the delocalized eigenvectors has to be obtained numerically, 
the fact that they are delocalized
allows one to determine their frequency $\omega$. Indeed, since the breather
is itself a localized state, delocalized Floquet eigenstates 
simply correspond to the above discussed linear EWs.
Their frequencies have been derived in Section II, and thus we can reconstruct
the delocalized part of the Floquet spectrum using them.

Stable breathers are characterized by all Floquet multipliers being
located inside or on the unit circle. 
Instabilities occur after collisions
of multipliers on the inner circle with radius (\ref{radius})
and a subsequent detaching from this circle towards larger absolute
values. Although in the dissipative case an 
additional change of the control 
parameter (dc bias $\gamma$) is necessary in order for  the 
corresponding Floquet eigenvalue(s) 
to cross and escape the unit circle,  
we may still look for collision conditions
and classify possible instability scenarios. 

Note that for Hamiltonian systems any collision and detaching from
the unit circle leads to an instability. Finite (even though possibly
weak) dissipation may drastically change the instability patterns
by selecting the strong instabilities (large detachments) over the
weak instabilities (small detachments). This is exactly what we
observe in our numerical studies. 

By applying the general stability analysis of nonlinear discrete systems 
\cite{via89} we obtain three possibilities for multiplier collisions. 
The first is realized when the collision takes place on the positive
real axis in the
complex plane. This implies that a multiplier is colliding with
its complex conjugate partner. With (\ref{floq}) it follows
that
\begin{eqnarray}\label{eq10}
\omega=m\Omega\;,
\end{eqnarray}
for any integer number $m$.
These are {\bf primary resonances} 
of the breather frequency or its
higher harmonics with any of the frequencies of the EEs.

The collision of
an eigenvalue with its complex conjugated partner can also take place on
the negative real axis
in the complex plane: 
\begin{eqnarray}\label{eq11}
\omega=(\frac{1}{2}+m)\Omega\;.
\end{eqnarray}
These are {\bf parametric resonances} of the breather state with  EEs .

The third case is realized when the collision takes place away from
the real axis.
Then a multiplier has to collide with a different
one, but not with its own complex conjugate partner. 
It follows that
\begin{eqnarray}\label{eq12}
\omega_{1}\pm\omega_{2}=m\Omega\;.
\end{eqnarray}
This is a {\bf combination resonance}, as the breather frequency (or its
multiple) has to match a sum (difference) of the 
frequencies of two different EE frequencies.

We stress again that EEs may be localized or delocalized. While we
may proceed with an analytical prediction of collisions using
delocalized Floquet multipliers, we have to use numerical calculations
to observe collisions involving localized ones. Especially the combination
resonance may involve either two delocalized, two localized or 
one delocalized and one localized Floquet eigenvalues. In addition
the above mentioned dissipation-induced selection of weak and strong
instabilities will result in some possible collisions being harmless (leaving
the breather stable)  while others will turn out to be important for
understanding breather instabilities.

\section{Numerical simulations of breather dynamics}

To study the breather dynamics, we performed  
direct numerical simulations of the set of equations
(\ref{2-8}). All simulations were carried out for JJLs with $N=10$ cells. 
We impose open boundary conditions and use the 4th order 
Runge-Kutta method. Time is measured in dimensionless time units. 
The initial value of the dc bias was $\gamma=0.8$. 
We choose proper initial conditions 
that lead to the relaxation of the system 
into
a particular breather state of left-right
symmetry with one resistive vertical junction, as in Fig.\ref{fig2}b . 
After a waiting time of 500 time units
we use the next 500 time units to calculate the time averaged characteristics
of the state. We then decrease the dc bias $\gamma$ by a tiny 
step of $\Delta\gamma=0.0005$ and repeat the 
procedure. We checked that our results do not change upon further
increase of the waiting time.
We varied the anisotropy $\eta$ and the inductance of the cell $\beta_L$
while the dissipation  $\alpha=0.1$ was fixed.
We will comment on hysteresis effects due to additional increasing
of the current in the conclusion.

\begin{figure}[htb]
\vspace{20pt}
\centerline{\psfig{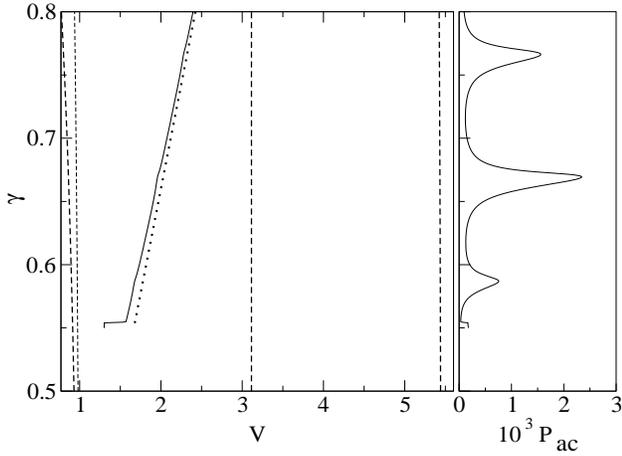}}
\vspace{2pt}
\caption{$I-V$ characteristics and the $P_{ac}$ dependence on the dc bias current
for $\alpha=0.1$, $\beta_L=0.2$, $\eta=1.15$.}
\label{fig5}
\end{figure}

\begin{figure}[htb]
\vspace{20pt}
\centerline{\psfig{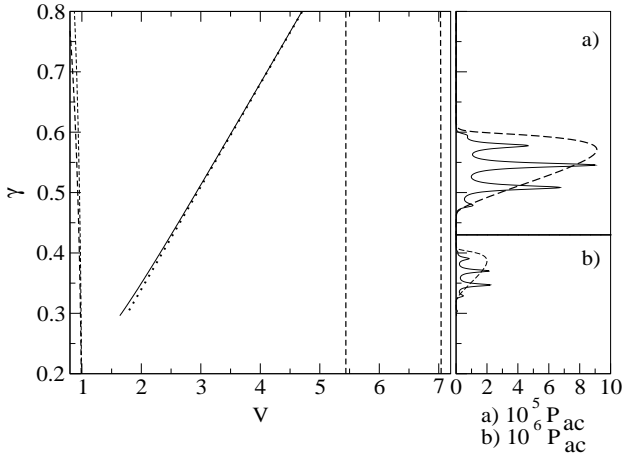}}
\vspace{2pt}
\caption{$I-V$ characteristics and the $P_{ac}$ dependence on the dc bias current
for $\alpha=0.1$, $\beta_L=0.2$, $\eta=0.35$.}
\label{fig6}
\end{figure}

\begin{figure}[htb]
\vspace{20pt}
\centerline{\psfig{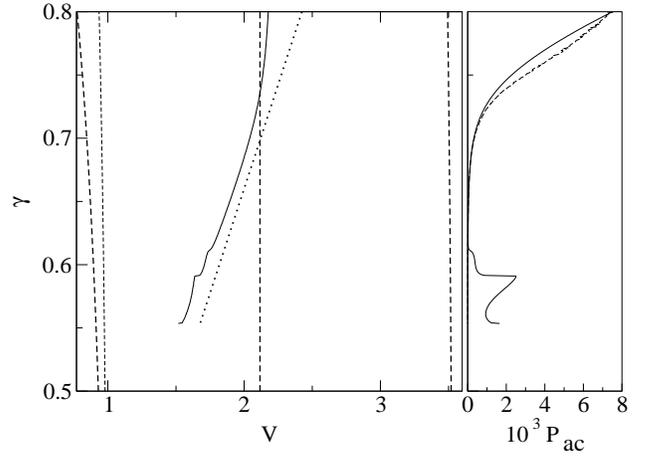}}
\vspace{2pt}
\caption{$I-V$ characteristics and the $P_{ac}$ dependence on the dc bias current
for $\alpha=0.1$, $\beta_L=0.5$, $\eta=1.15$.}
\vspace{20pt}
\label{fig7}
\end{figure}

\begin{figure}[htb]
\vspace{20pt}
\centerline{\psfig{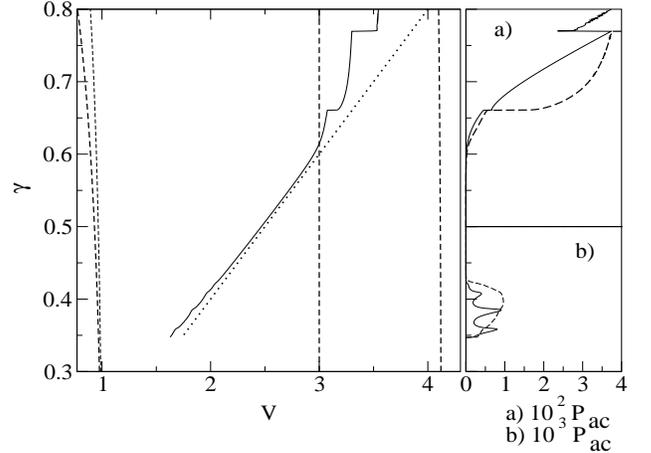}}
\vspace{2pt}
\caption{$I-V$ characteristics and the $P_{ac}$ dependence on the dc bias current
for $\alpha=0.1$, $\beta_L=0.5$, $\eta=0.5$.}
\label{fig8}
\end{figure}

\begin{figure}[htb]
\vspace{20pt}
\centerline{\psfig{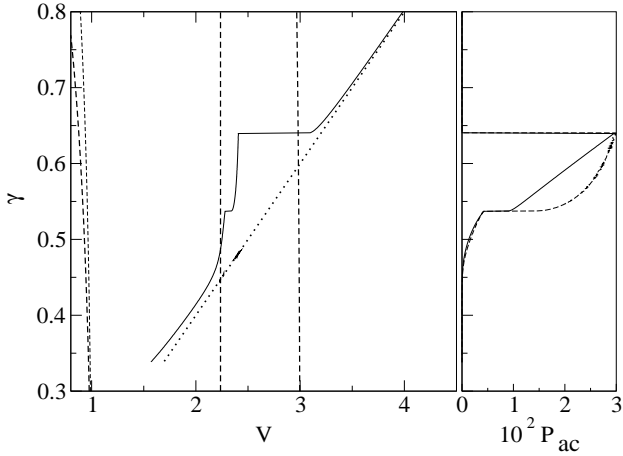}}
\vspace{2pt}
\caption{$I-V$ characteristics and the $P_{ac}$ dependence on the dc bias current
for $\alpha=0.1$, $\beta_L=1.0$, $\eta=0.5$.}
\vspace{20pt}
\label{fig9}
\end{figure}

\begin{figure}[htb]
\vspace{20pt}
\centerline{\psfig{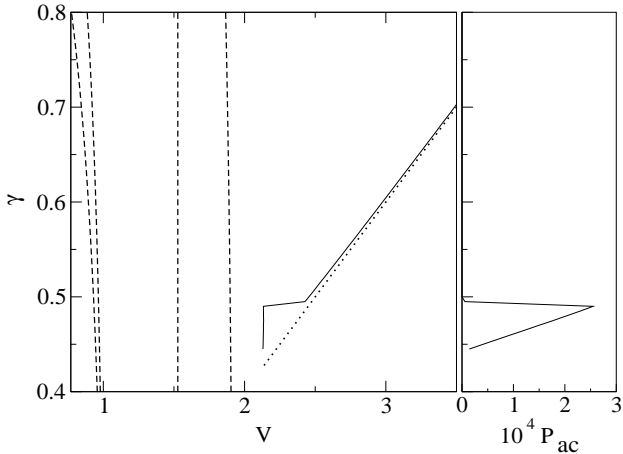}}
\vspace{2pt}
\caption{$I-V$ characteristics and the $P_{ac}$ dependence on the dc bias current
for $\alpha=0.1$, $\beta_L=3.0$, $\eta=0.5$.}
\label{fig10}
\end{figure}

\begin{figure}[htb]
\vspace{20pt}
\centerline{\psfig{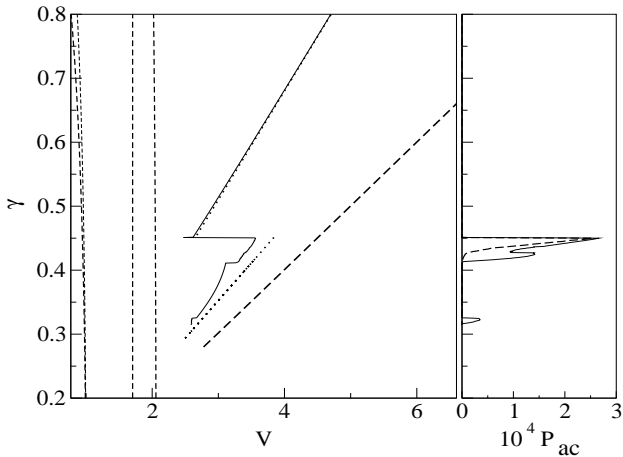}}
\vspace{2pt}
\caption{$I-V$ characteristics and the $P_{ac}$ dependence on the dc bias current
for $\alpha=0.1$, $\beta_L=3.0$, $\eta=0.35$.}
\label{fig11}
\end{figure}

There are three different ways to monitor the simulations.
The first one is the $I-V$
characteristics, that is the dependence of the averaged
voltage drop across the resistive vertical junction on the dc bias. 
Furthermore we obtain the power $P_{ac}$ (\ref{pac}) of ac oscillations of the vertical junction 
at the edge of JJLs.
Finally we generate time-resolved images (movies) of the full dynamical
behavior of the ladder so that we can visually
check whether the
system still resides in the initially chosen breather state, or switches
into another state.
Our results are presented in Figs. \ref{fig5}-\ref{fig11},
where each figure consists of two parts. In the left hand parts the
$I-V$ characteristics are shown (solid lines) together with the
approximate results from section III (dotted lines).
The vertically oriented dashed lines indicate the
band edges of the linear EWs. In the right hand part of the
figures we show the dependence of $P_{ac}$ (solid lines) 
on the dc bias together
with our approximate analysis from section IV and Appendix A (dashed lines)
where appropriate.

We start with the case
of small $\beta_L$ values. For
$\beta_L=0.2$ and $\eta=1.15$ (Fig.\ref{fig5}) the breather is easily excited,
and its frequency is located below $\omega_+(q)$. In the $P_{ac}$ plot
we observe peaks that are due to the resonance of the second 
and third harmonics
of the breather with $\omega_+(q)$.  These resonances are primary ones
as discussed above. Note that their presence
is barely seen on the $I-V$ curve. The series of observed peaks
is related to the finite size of the system, and therefore 
to the resonant interaction of the breather with a discrete set of 
cavity modes 
as was discussed at the 
end of section II. 
We tested our interpretation by 
increasing the size of the system and observed the predicted increase
in the number of resonance peaks.  Close to the lowest possible
current (around $\gamma=0.55$) we observe a switching to 
another breather state,
which however has the same symmetry and spatial structure. Note that
shortly after this switching (upon further lowering of the current)
we lose the breather and the system switches 
to the superconducting ground state. 

For lower values of the 
anisotropy $\eta=0.35$ (Fig.\ref{fig6}) 
the resonances are again not detectable in
the $I-V$ curve. However, by monitoring $P_{ac}$ 
we observe the singularities that correspond
to the primary resonance $ 2\Omega=\omega_+(q)$ . 
Moreover, at the dc bias $\gamma\approx 0.35$ we  detect a weak third-order
primary resonance $3\Omega=\omega_{+}(q)$ in the breather tail. 
The dashed line in the right part of the figure
is the prediction of $P_{ac}$ using our approximate 
tail analysis. Note that our approximate tail analysis 
is based on the assumption 
of a dense spectrum of EWs. Consequently the calculated $P_{ac}$ presents an
envelope of the numerically observed series of discrete peaks.

%

%

Next we increased the inductance of the cell to $\beta_L=0.5$. In 
Fig.\ref{fig7} we show the results for $\eta=1.15$. For the initial 
value of the dc bias $\gamma~=~0.8$
the breather frequency is already located inside the $\omega_+(q)$ band 
of EWs, and this primary resonance is observed in the $I-V$ curve. 
Indeed, the slope of the $I-V$ curve is larger than the 
prediction (\ref{3-7}) which does not take into account resonant
interactions with EEs.
With decreasing dc bias, the breather frequency is lowered and
the above primary resonance disappears. However at lower current values
the next primary resonances $2\Omega=\omega_+(q)$ occur and are
observable, both in the breather tail and in the $I-V$ characteristics.

The primary resonance structures ($\Omega~=~\omega_+(q)$)
in the large current domain 
are also observed 
for smaller values of the anisotropy parameter $\eta=0.5$ 
(Fig.\ref{fig8}).
In this case, they
manifest themselves through {\it resonant steps} 
in the $I-V$ curve. \cite{comment1}
At lower values of the dc bias we again observe primary resonances 
with $m~=~2$.
%

%

For
$\beta_L=1.0$ and $\eta=0.5$ (Fig.\ref{fig9}) the breather frequency
is located above $\omega_+(q)$ for large current values.
Upon decreasing the dc bias we observe a peculiar switch to a different
breather state with the same spatial structure but a lower frequency 
located inside $\omega_+(q)$. The most interesting feature here
is that shortly before the switching the breather frequency is clearly
larger and outside of the $\omega_+(q)$ region. 
Upon further lowering of the dc bias we observe primary resonances 
$\Omega=\omega_+(q)$, and
corresponding resonant steps in $I-V$ curve.

%

Let us increase the inductance of the cell $\beta_L$ even further. 
For $\beta_L=3$ and $\eta=0.5$ (Fig.\ref{fig10})
we again find that breather
frequencies are located above $\omega_+$ for large current values. 
Similar to the previous case we observe 
a switching when the breather frequency is clearly outside (above)
the branch $\omega_+(q)$. This brings the system into
another breather state with the same spatial
structure, but with a frequency again located above $\omega_+$. 
This highly
nonlinear state is then lost by switching to the superconducting
ground state after further decrease of the dc bias. 

%

Now we come to an interesting observation.
Lowering the anisotropy $\eta=0.35$ (Fig.\ref{fig11})
we again observe the switching at a breather frequency
being located above $\omega_+(q)$. However the switching
{\it increases} the voltage drop. The new state is of different
internal structure. We remind the reader that all previous numerical
results have been obtained for a breather with a structure as in
Fig.\ref{fig2}b. Here we find that after the switch 
the new breather state is
characterized by {\it three} vertical junctions 
being in the resistive state. At the same time the
symmetry is broken. In fact  
the new
state exactly corresponds to the example given in Fig.\ref{fig2}d.
Note that similar switchings (which lead to
an increase of the number of resistive junctions)
have been reported in early experimental
studies. \cite{pbdaavusfyz00,comment2}
The left-right symmetry breaking  
leads to new interesting features in the breather tails.
The no-symmetry breather has two times lower frequency
than the voltage drop across the vertical junction. Thus the new
breather frequency is inside the upper band
$\omega_+(q)$, and a primary resonance is clearly observed in
the $P_{ac}$ dependence on $\gamma$. 

In order to test the influence of small fluctuations on the $I-V$ curves 
we repeated the simulations in the presence of small noise with amplitude 
$\approx 10^{-8}$. All obtained results are stable {\it except}
the switching outcome in Fig.\ref{fig11} . While this switching occurs
at the same current value, the {\it new} breather state is changed. In
particular we observed the left-right symmetry breather with three junctions
being in the resistive state. 
Thus we find extreme sensitiveness of the outcoming breather
structure (including its symmetries) to small fluctuations. This implies
that the boundaries of the volumes of attraction of different (breather)
attractors are entangled in a very peculiar way. We
can reliably predict the switching position, but not the outcome
of the switching.

We also numerically simulated the breather dynamics in the JJL with 
an extremely large inductance of the cell $\beta_L=500$. We did not find 
any indication
of resonances and instabilities. We argue that the reason for that 
is the weak dispersion of
the linear EWs for such large values of $\beta_L$. 
This implies that interactions
along the ladder are weak. 
The breather is continued to small
current values until it switches to the superconducting state at the dc 
dc bias $\gamma~=~0.22$ for $\eta=0.35$. 
This particular value can be obtained by making use of the simple dc analysis 
(\ref{vlr}) and the standard theory of the retrapping current 
in a single small 
Josephson 
junction \cite{etjjmabtpo00,abgp82}
\begin{equation}\label{ret}
\gamma_{r} = (1+2\eta)\frac{4\alpha}{\pi}\;.
\end{equation}
This equation yields a value of 0.22 for the considered case, in good agreement
with the numerical observation. Note that within this theory retrapping
occurs purely due to energy considerations, not due to resonances
(or instabilities). 

It is very important to notice 
that for cases with small or intermediate values of inductance
of the cell (Fig.\ref{fig5}-\ref{fig11})  the observed 
currents at which we lose the breather state and switch to the
superconducting one
exceed the expected retrapping values (\ref{ret}). 
We will explain this disagreement in the next section.

Motivated by the above findings 
we investigated the loss of the HWS upon lowering the current.
We recall that in this state {\it all} vertical junctions are
resistive and {\it all} horizontal ones are superconducting.
Usually it is assumed that the HWS loss is again due to a 
standard retrapping mechanism.
It is important that any numerical simulation of such a process
is done with the addition of some weak noise, because the 
processor will otherwise perform a perfect simulation of a single
junction repeated $N+1$ times. We chose $\beta_L=3$ and $\eta=0.35$.
The $I-V$ characteristic is shown by a thick dashed line in Fig.\ref{fig11} .
The expected retrapping current $4\alpha/\pi=0.127$ is clearly
not reached. Instead we observe the loss of the HWS at $\gamma=0.273$.
At the same time it follows from equations (\ref{2-8}) that
the HWS exists as a solution down to the retrapping current of a 
single junction, i.e. down to $\gamma=0.127$ ! So in this case we
conclude that the numerically observed loss of the HWS at $\gamma=0.273$
is {\it due to an instability}. The HWS continues to
exist as a solution down to the standard retrapping current,
but it is an {\it unstable} state. This novel result is very important,
since very often the current value of HWS loss in the absence
of a magnetic field is used to estimate 
different parameters of the system {\it assuming} that the HWS
is behaving similar to a single junction. Our results show that this
is definitely {\it not} the case.

%

\section{Evaluation of resonances and EWs spectroscopy}

This section is devoted to a quantitative explanation of the
observed resonances and switchings. 

\subsection{Primary resonances}

Primary resonances are characterized by $m\Omega=\omega$ 
where $\omega$ is some EE frequency.
We detected various primary resonances with {\it extended} EWs.
The case $m=1$,
which corresponds to the breather frequency being located inside
the $\omega_+(q)$ band, shows up with resonant steps in the 
$I-V$ curves (see, Figs. \ref{fig8}-\ref{fig9}).
The finite number of observed resonant steps is
due to the discrete spectrum of the excited cavity modes.
In addition we observe strong variations of the breather tail amplitudes.

Higher order primary resonances ($m=2,3$) are much less pronounced
in the $I-V$ characteristics. They mainly lead to a weaker localization
of the breather tail and can be clearly detected in the form of sharp peaks 
in the $P_{ac}(\gamma)$ dependence.
Since the breather in our case has left-right symmetry, the
only linear cavity modes which can be exited are symmetric 
(see section II). These 
modes are characterized by even values of $k$ (\ref{obc}).
We start our evaluation of these resonances
with the case shown in Fig.\ref{fig6}.
We determine the $\gamma$ value of each observed peak in $P_{ac}(\gamma)$
and thus obtain the corresponding breather frequency $\Omega(\gamma)$.
We then compare its multiples with the discrete spectrum of 
linear mode frequencies of the $\omega_+$ branch. The numbers are
listed in Table I. We find that all observed resonances are
due to symmetric linear modes (even $k=2,4,6,8,10$) as expected.
\begin{table}\label{tab1}
\begin{tabular}{|c||c|c|c|}
\hline
\itshape $\hspace*{5mm} k \hspace*{5mm}$&\itshape  $
\hspace*{5mm} \omega_{+}(q_k)
\hspace*{5mm}$&\itshape $ \hspace*{5mm} 2\Omega
\hspace*{5mm}$&\itshape
\hspace*{5mm} $3\Omega$
\hspace*{5mm}\\
\hline\hline
$1$ &5.475&&\\
$2$ &5.581&5.594&5.590\\
$3$ &5.745&&\\
$4$ &5.949&5.947&5.945\\
$5$ &6.174&&\\
$6$ &6.400&6.387&6.384\\
$7$ &6.609&&\\
$8$ &6.788&6.769&6.772\\
$9$&6.923&&\\
$10$&7.008&6.960&6.958\\
\hline
\end{tabular}
\caption{Comparison of theoretical predictions of primary
resonances with the numerical results from Fig.\ref{fig6}. 
First column: $k$. 
Second column: the spectrum of EWs $\omega_{+}(q_k)$ 
(cf. (\ref{obc}).
Third column: $2\Omega$ obtained from the peak positions of
$P_{ac}(\gamma)$ in Fig.\ref{fig6}(a).
Fourth column: $3\Omega$ obtained from the peak positions of
$P_{ac}(\gamma)$ in Fig.\ref{fig6}(b).}
\end{table}

The same method of analysis allows us to conclude that the 
three peaks in Fig.\ref{fig5} (in decreasing order of dc bias)
are due to the following
resonances: $(m,k)=(2,6);(2,4);(3,8)$.
Similarly the shoulder and the peak in Fig.\ref{fig6} are due to
resonances with $(m,k)=(2,10);(2,8)$. 
Finally the resonances in Fig.\ref{fig8} correspond to the values
$(m,k)=(2,8);(2,6);(2,4)$. Note that in all of these cases
the deviations between the theoretical and observed numbers are much
less than the frequency difference between adjacent cavity modes.

\subsection{Parametric resonances}

So far we did not comment on the nature of the switching from a breather
state to the superconducting state for small and intermediate
values of $\beta_L$. The Floquet analysis results show that 
{\it all} these switchings are due to an {\it instability}
of the breather. In terms of Floquet multipliers,
all of these instabilities are due to a collision of two {\it localized}
Floquet multipliers on the negative real axis. 
The breather state continues to exist as a solution to the dynamical
equations for lower current values, but it is unstable. 
Note that the so-called retrapping mechanism instead 
(as for a single junction) uses the critical {\it current} value
as a criterion for retrapping. This argument is based purely on
energy considerations and does not take into account any
resonance mechanism. This is not surprising, as a single junction
has no other degrees of freedom it may resonate with. Below
the retrapping current, the resistive state disappears in this case.
So we may state that the switching from a breather state
to the superconducting one as observed in our simulations
is usually driven by {\it resonances} with localized EEs
(frequency matching) and is {\it not} due to energy effects
(current value matching). 

\subsection{Combination resonances}

Let us discuss the nature of the switchings of the breather
for intermediate $\beta_L$ values when the breather frequency
is located {\it above} the branch $\omega_+(q)$. These switchings
are again due to an instability. It is characterized by
Floquet multipliers colliding away from the real axis. 
As discussed in Section V this corresponds to a combination resonance.
The numerical Floquet analysis shows that one of the two participating
multipliers is a {\it localized} one (which bifurcates from the
lower branch $\omega_-(q)$)., while the second one belongs
to the delocalized spectrum of $\omega_+(q)$. The Floquet multiplier
which finally leaves the unit circle is a {\it localized} one.
So again the instability of the breather is driven by a localized
perturbation. 
%
\begin{figure}[htb]
\vspace{20pt}
\centerline{\psfig{figure=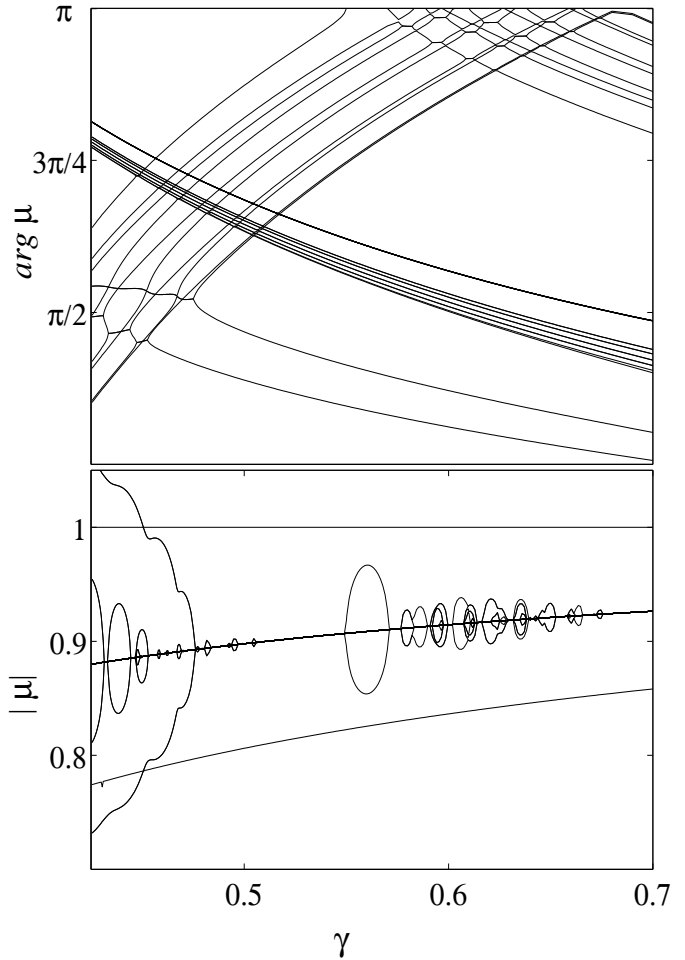,width=90mm,height=130mm}}
\vspace{2pt}
\caption{Arguments and absolute values of Floquet multipliers versus
$\gamma$ for the breather state in Fig.\ref{fig11}.}
\label{fig12}
\end{figure}
In Fig.\ref{fig12} the dependence of the arguments
and absolute values of all relevant Floquet multipliers is shown for the
breather of Fig.\ref{fig11}. 
For convenience we do not plot the complex conjugate multipliers
and restrict the arguments to $0 \le arg(\mu) \le \pi$. The narrow
band $\omega_-(q)$ and broad band $\omega_+(q)$ are nicely observed.
The degenerate band $\omega_0$ is
located slightly above $\omega_-(q)$. 
This band does not interact with other multipliers when crossing
them, as expected from our analytical considerations.
The two separated arguments which are located {\it below}
the $\omega_-(q)$ band have {\it localized} eigenvectors. \cite{locmodenote}

In the plot of the absolute values we observe the predicted
values $\mu=1$ and $e^{-\alpha T_b}$. The multipliers which
correspond to lines between these two states generally reside
on the circle with radius (\ref{radius}). Many of
them depart from this circle due to collisions. 
At current values
of $0.55 < \gamma < 0.7$ we observe parametric resonances 
$2\Omega = \omega_+(q)$, which belong to the set of weak resonances and do
not evolve into a global instability. However it is possible
that a slight variation of control parameters (e.g. decreasing
the damping $\alpha$) might change these resonances into strong ones.
Then we would expect sudden instabilities of the breather state at these
large current values. In our case the global instability is realized
when one of the localized multipliers collides with the $\omega_+(q)$
band around $\gamma =0.48$. Subsequent lowering of the current
leads to a fast escape of this multiplier from the unit circle and to
the observed switching.  

The importance of localized EEs for the destabilization of
a breather is simply due to the localized nature of the latter.
It is hard (if not impossible) for a breather to generate
a parametric instability through extended EWs alone, as 
these excitations are damped out far from the breather center.
In contrast localized EEs do not travel away from the breather
center. These modes can be effectively excited by the breather,
leading to an instability of the latter.  

\begin{figure}[htb]
\vspace{20pt}
\centerline{\psfig{figure=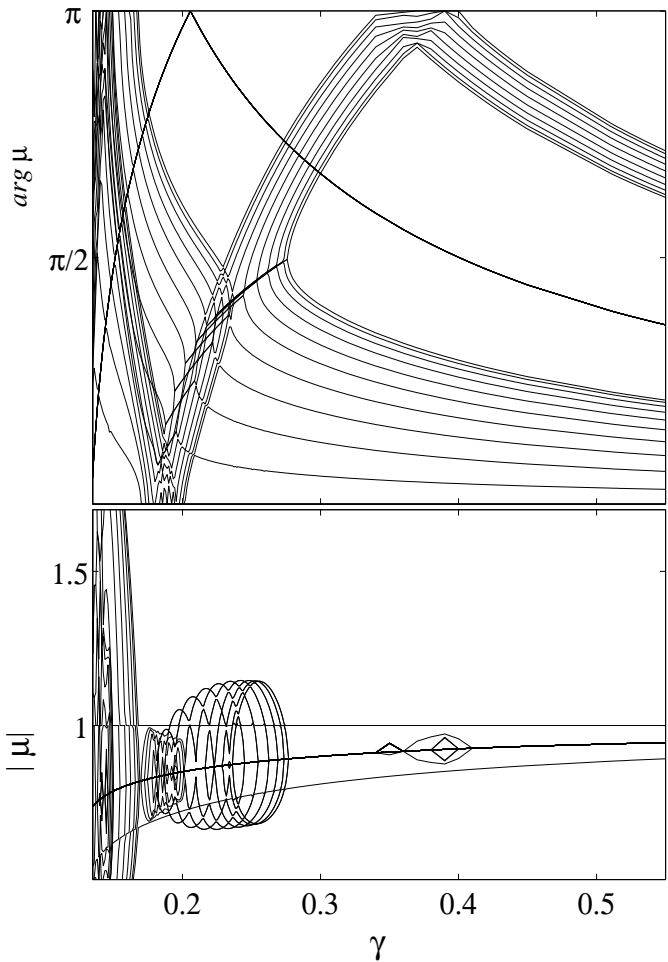,width=90mm,height=130mm}}
\vspace{2pt}
\caption{Arguments and absolute values of Floquet multipliers versus
$\gamma$ for the HWS in Fig.\ref{fig11}.}
\label{fig13}
\end{figure}

To understand the nature of the observed instability of the HWS
we show a similar Floquet multiplier plot in Fig.\ref{fig13}.
%
As in the previous plot, we observe weak parametric resonances
of the upper EW band at current values $\gamma \sim 0.38$ which
do not evolve into a global instability. 
Again the observed instability is driven by a combination resonance
at $\gamma=0.175$.
Since the HWS is an extended state, all Floquet multipliers are 
also extended. The combination resonance is due to the collision
of two Floquet multipliers belonging to the two EW branches
$\omega_{\pm}(q)$ with $q=\pi$ (we remind the reader that
the EW spectrum of the HWS is different from that of the superconducting
state and can be obtained by putting $\gamma=1$ in 
(\ref{2-13})). Note that indeed for the
present case, the frequency of the HWS at the instability equals
$2.73$, while the value of the combination 
$\omega_-(\pi) + \omega_+(\pi) = 2.69$.
For current values $\gamma \sim 0.18$ the HWS is becoming
{\it stable} again. However around $\gamma = 0.17$ another
even stronger instability due to parametric resonance sets in,
which brings the HWS to the next instability well above
the expected retrapping current.

\section{Conclusion}

We have presented analytical and numerical studies of breather properties
in Josephson junction ladders. Our results confirm and substantially extend
early suggestions that breathers may resonate in different ways
with localized and extended electromagnetic modes. 
The numerical studies have been done
in a parallel manner to the  way  experiments are conducted. The variation
of the control parameter $\gamma$ allows one to continuously
change the breather frequency, whereas the linear mode spectrum is not
significantly changed in the domain of interest.
We observed primary resonances $m\Omega=\omega_+(q_k)$ with extended
EWs,
parametric resonances $\Omega=2\omega$ with localized EEs,
and combination resonances $m\Omega = \omega+\omega_+(q_1)$
with a localized EE and a delocalized EW participating. 
We also observed a combination resonance which leads to
a switching from a small breather (one resistive vertical junction)
to a larger one (three resistive vertical junctions) together with
a possible symmetry lowering of the breather.

The primary resonances with extended EWs lead to singularities
in the breather tails. 
This allows one to develop a spectroscopy of EWs by monitoring
$P_{ac}$ versus $\gamma$. Such a spectroscopy 
can be experimentally realized, e.g. with the help of
a well known Josephson junction detector technique. \cite{kkl86,Benz,Ustinov}
It could be important for obtaining
a coherent source of high frequency radiation,
since in such a resonance the whole breather tail starts to coherently
oscillate with large amplitudes. The resistive breather center
serves as a region of energy input via a dc bias.

Our studies show that the main control parameter (in addition to
the dc bias) is the self inductance $\beta_L$. For small $\beta_L$
values the breather frequency is located between the two 
branches of EWs, $\omega_{\pm}(q)$. One may perform spectroscopy of EWs
of the upper branch, or observe parametric instability of a breather due
to localized modes. 
Moreover as $\beta_L$ increases, resonant steps in the $I-V$ characteristics
can be observed.
For intermediate $\beta_L > 1$ values, the 
breather frequency is located above $\omega_+(q)$. In this case
we observe combination resonances due to localized and delocalized EEs,
which may result in a unusual sensitivity of the switching outcome
on small fluctuations. 
Large $\beta_L$ values stabilize the breather
states, make resonances impossible and lead to a standard retrapping
mechanism for breather switching to the superconducting ground state.
This is likely the situation for the reported experimental data in
[Ref.~\onlinecite{pbdaavu00}].
However lower $\beta_L$ values allow for the appearance of the above listed
resonances (and perhaps even other still unobserved resonances). We believe that
our findings will help to make the proper parameter choice when 
designing new ladders for experiments.

Note that throughout our studies we always {\it decrease} the dc bias $\gamma$.
Let us consider a breather state $Nr.1$ which becomes unstable
upon lowering the dc bias at a certain value $\gamma_1$.
Let this be a case where the system will switch
to another
stable breather state $Nr.2$. This new breather state is in fact
keeping its stability not only upon further lowering of
the dc bias, but also upon a reversing (increasing) of the dc bias.
Thus we find that there exist dc bias windows
in which both the starting breather state $Nr.1$ and the 
new breather state $Nr.2$
are stable. Even though their dc spatial structures (cf. Fig.\ref{fig2})
may be identical, the average voltage drops (and frequencies) are in
general different. Further increasing of the dc bias while staying
on the new breather state $Nr.2$ will lead to an instability and switching
at $\gamma_2 > \gamma_1$.
In case the switching brings the system back to the breather state
$Nr.1$ we are faced with the well-known two-state hysteresis phenomenon in
Josephson junction systems. However we also observed cases when
the switching due to an instability of breather state $Nr.2$ (upon 
increasing the dc bias) brings the system to yet {\it another} state, which
differs from breather state $Nr.1$, e.g. simply to the HWS. In such a case
the hysteretic behavior is of more complex nature. To keep the
discussion of our studies as clear as possible we did not
present data for {\it increasing} current.

Another observation is that the expected values of the retrapping
current based on a pure dc analysis \cite{etjjmabtpo00} 
are too low to match the observed values at which the
breather switches to the superconducting ground state. Only for
very large $\beta_L$ values do we observe agreement. For all other
cases the breather switches to the superconducting state via
an instability driven by parametric resonance with localized EEs.
Moreover the HWS also undergoes an instability which is due
to combination resonances with extended EWs.

In this work we always started with a breather configuration as in
Fig.\ref{fig2}b,
at large current values. It may be expected that the results for
other starting configurations will show up with similar
properties, and perhaps with additional types of resonances as well.
This may be due to the fact that the structure of the phase space 
is very complex, being separated in many different regions of attraction
of different attractors. It is this complexity which
makes the understanding of breather properties both a
fascinating and complicated undertaking.
\\
\\
\\
{\bf Acknowledgements}
\\
\\
We thank A. Benabdallah, P. Binder, M. Schuster and A. V. Ustinov for
valuable discussions and F. V. Kusmartsev and J. J. Mazo
for sending us their preprints prior publication.
This work was supported by the Deutsche Forschungsgemeinschaft and
by the European Union under the RTN
project
LOCNET HPRN-CT-1999-00163.

\appendix
\section{Power of ac librations at the edge of a JJL}

Here we derive the time-average power of ac librations at the edge of a JJL.
This characteristics is proportional to the average 
kinetic energy $\langle \dot{\phi}^2/2\rangle$.
In order to obtain an expression for the kinetic energy,  
we have to determine the dynamics of the junction at the JJL edge.
For this we write the system of equations (\ref{4-4}) in a matrix form:
\begin{eqnarray}\label{a-1}
\hat{A}\vec{v}=0\;.
\end{eqnarray} 
Here, $\vec{v}$ is an unknown vector 
\begin{eqnarray}
\vec{v}=\left(\begin{array}{c}\Delta_v\\\Delta_h\end{array}\right)\;,
\end{eqnarray}
and $\hat{A}$ is a $2\times 2$ matrix 
\begin{eqnarray}
\hat{A}(\lambda)=\left(\begin{array}{cc}a_{11}(\lambda) & a_{12}(\lambda)\\ a_{21}(\lambda) &
a_{22}(\lambda)\end{array}\right)~,
\end{eqnarray}
where $a_{11}(\lambda)~=~A-\frac{2}{\beta_L}\cosh \lambda$, 
$a_{12}(\lambda)~=-\frac{2}{\beta_L}(1-e^{-\lambda})$, 
$a_{21}(\lambda)~=-\frac{1}{\beta_L \eta}(1-e^{\lambda})$, and
$a_{22}(\lambda)=B$. 
A nonzero solution exists if the determinant of $\hat{A}$ vanishes: 
$\det\hat{A}(\lambda_0)=0$.
The parameter $\lambda_0$ was determined in section IV (\ref{4-7}).

The components of the vector $\vec{v}$ satisfy the condition:
\begin{eqnarray}\label{a-2}
\Delta_v=-\frac{a_{22}}{a_{21}}\Delta_h\;.
\end{eqnarray}
To determine the components $V$ and $H$ separately we have to impose an 
additional condition at the breather center. 
This condition is not 
known exactly due to the complex dynamics in the resistive breather center.
Nevertheless 
the ac librations in the breather tails are weakly depending on it.
Here we use the simplest normalization condition 
\begin{eqnarray}
|\Delta_v|^2+|\Delta_h|^2=1\;.
\end{eqnarray}

Substituting (\ref{a-2}) into (A5) we obtain
\begin{eqnarray}
\Delta_h&=&\frac{|a_{21}|^2}{\sqrt{|a_{21}|^2+|a_{22}|^2}}\nonumber\\
\Delta_v&=&-\frac{a_{22}\overline{a}_{21}}{\sqrt{|a_{21}|^4+|a_{22}|^2|a_{21}|^2}}\;.
\end{eqnarray}
Due to the up-down symmetry in the breather tail, the dynamics of 
Josephson phases at the edge of the JJL can be written in the form
\begin{eqnarray}
\varphi_n^v & = &
-\frac{\Re(a_{22}\overline{a}_{21})e^{\Re(\lambda_0)n}\cos(\Im(\lambda_0)n+\Omega
t)}{\sqrt{|a_{21}|^4+|a_{22}|^2|a_{21}|^2}}\nonumber\\
&&+\frac{\Im(a_{22}\overline{a}_{21})e^{\Re(\lambda_0)n}\sin(\Im(\lambda_0)n+\Omega
t)}{\sqrt{|a_{21}|^4+|a_{22}|^2|a_{21}|^2}}\;,\\
\varphi_n^h&=&\frac{|a_{21}|^2e^{\Re(\lambda_0)n}\cos(\Im(\lambda_0)n+\Omega
t)}{\sqrt{|a_{21}|^2+|a_{22}|^2}}\nonumber\;.
\end{eqnarray}

We finally obtain the expression for the average kinetic energy (for $n<0$)
\begin{eqnarray}
\frac{1}{2}\left<\dot{\varphi}_n^{v^2}\right>&=&\frac{\Omega^2|a_{22}\overline{a}_{21}|^2
e^{\Re(\lambda_0)2n}}{4(|a_{21}|^4+|a_{22}|^2|a_{21}|^2)}\;.
\end{eqnarray}

\section{Linear stability of the breather in JJL.}
The stability of periodic motion is analyzed with the help of 
the Floquet theory. 
\cite{via89,swhsjzshstpo96} 
Linearizing the system (\ref{2-8}) around a 
time periodic breather 
solution, we obtain  
\begin{eqnarray}
   \ddot{\epsilon}_{n}^v+\alpha\dot{\epsilon}^v_{n}+A_n^v(t)\epsilon_{n}^v 
   &=&\frac{1}{\beta_{L}}(\triangle\epsilon_{n}^v+\nabla\epsilon_{n-1}^h-
   \nabla\tilde{\epsilon}_{n-1}^h)\nonumber \\
   \ddot{\epsilon}_{n}^h+\alpha\dot{\epsilon}_{n}^h+A_n^h(t)\epsilon_{n}^h 
   &=&-\frac{1}{\eta\beta_{L}}(\nabla\epsilon_{n}^v+\epsilon_{n}^h-
   \tilde{\epsilon}_{n}^h) \\
   \ddot{\tilde{\epsilon}}{}_{n}^h+\alpha\dot{\tilde{\epsilon}}{}_{n}^h+
   \tilde{A}_n^h(t)\tilde{\epsilon}_n^h &=&\frac{1}{\eta\beta_{L}}(
   \nabla\epsilon_{n}^v+\delta_{n}^h-\tilde{\epsilon}_{n}^h)\nonumber\;, 
\end{eqnarray}
where $A_n(t)$ are time-periodic coefficients determined by the
given breather state.

The substitution 
\begin{eqnarray}
\left(\begin{array}{c}\epsilon_n^v \\ \epsilon_n^h \\
\tilde{\epsilon}_n^h\end{array}\right)=e^{-\frac{1}{2}\alpha
t}\left(\begin{array}{c}\kappa_n^v \\ \kappa_n^h \\
\tilde{\kappa}_n^h\end{array}\right)
\end{eqnarray}
allows one to eliminate the dissipation
\begin{eqnarray}  
\ddot{\kappa}_{n}^v+B_n^v(t)\kappa_{n}^v&=&\frac{1}{\beta_{L}}(\triangle\kappa_{n}^v+\nabla\kappa_{n-1}^h-\nabla\tilde{\kappa}_{n-1}^h)\nonumber \\
   \ddot{\kappa}_{n}^h+B_n^h(t)\kappa_{n}^h  
&=&-\frac{1}{\eta\beta_{L}}(\nabla\kappa_{n}^v+\kappa_{n}^h-\tilde{\kappa}_{n}^h) \\
\ddot{\tilde{\kappa}}{}_{n}^h+\tilde{B}_n^h(t)\tilde{\kappa}_n^h&=&\frac{1}{\eta\beta_{L}}(\nabla\kappa_{n}^v+\kappa_{n}^h-\tilde{\kappa}_{n}^h)\nonumber\;, \\
\end{eqnarray}
where $B_n(t)=-\frac{1}{4}\alpha^2+A_n(t)$.
Introducing new variables
\begin{eqnarray}
z_n^v=\kappa_n^v\;\;,\;\; z_n^h=\sqrt{\eta}\kappa_n^h\;\;,\;\; 
\tilde{z}_n^h=\sqrt{\eta}\tilde{\kappa}_n^h\;,
\end{eqnarray}
we find the system of equations
\begin{eqnarray}
\ddot{z}_n^v+B_n^v(t)z_n^v & =& \frac{1}{\beta_L}\triangle
z_n^v+\frac{1}{\beta_L\sqrt{\eta}}(\nabla z_{n-1}^h-\nabla\tilde{z}_{n-1}^h)\nonumber\\
\ddot{z}_n^h+B_n^h(t)z_n^h & = &-\frac{1}{\eta\beta_L}(z_n^h-\tilde{z}_n^h)-\frac{1}{\beta_L\sqrt{\eta}}\nabla
z_n^v\\
\ddot{\tilde{z}}_n^h+\tilde{B}_n^h(t)\tilde{z}_n^h & = &
\frac{1}{\eta\beta_L}(z_n^h-\tilde{z}_n^h)+\frac{1}{\beta_L\sqrt{\eta}}\nabla
z_n^v\nonumber\;.
\end{eqnarray}
These equations describe a {\it Hamiltonian} system, namely
\begin{eqnarray}
\dot{\vec{z}}_n&=&\frac{\partial{\cal H}}{\partial \vec{p}_n}\nonumber\\
\dot{\vec{p}}_n&=&-\frac{\partial{\cal H}}{\partial \vec{z}_n}\;,
\end{eqnarray}
where $\vec{z}_n=(z_n^v,z_n^h,\tilde{z}_n^h)\;,\;\vec{p}_n=(p_n^v,p_n^h,\tilde{p}_n^h)$ 
and the Hamiltonian ${\cal H}(\vec{z}_n,\vec{p}_n,t)$ is
\begin{eqnarray}
{\cal H} & = & \frac{1}{2}\sum\limits_n [
p_n^{v^2}+p_n^{h^2}+\tilde{p}_n^{h^2}]\nonumber\\
& & +\frac{1}{2}\sum\limits_n [
B_n^v z_n^{v^2}+B_n^h z_n^{h^2}+\tilde{B}_n^h\tilde{z}_n^{h^2}]\nonumber\\
& & +\frac{1}{2\beta_L}\sum\limits_n(z_n^v-z_{n-1}^v)^2\\
& & +\frac{1}{2\beta_L\eta}\sum\limits_n(z_n^h-\tilde{z}_{n}^h)^2\nonumber\\
& & +\frac{1}{\beta\sqrt{\eta}}\sum\limits_n z_n^v [
z_{n-1}^h-\tilde{z}_{n-1}^h-z_n^h+\tilde{z}_n^h]\nonumber\;.
\end{eqnarray}
Since the particular Hamiltonian can be represented in a general quadratic form, 
the symplectic product of two different trajectories
$\{\vec{p}_n(t),\vec{z}_n(t)\}$ and  $\{{\vec{p}_n}^{\hspace{0.7mm}\prime}(t),\vec{z}_n^{\hspace{0.7mm}\prime}(t)\}$
does not change in time \cite{via89}
\begin{eqnarray}\label{sympprod}
{\cal I}=\sum\limits_n [\vec{p}_n^{\hspace{0.7mm}\prime} (t)\vec{z}_n(t)-\vec{p}_n(t)\vec{z}_n^{\hspace{0.7mm}\prime}(t)]\;.
\end{eqnarray}

Rewriting our set of equations in the form
\begin{eqnarray}\label{eq14}
\delta\dot{\vec{z}}_n&=&\frac{\partial^2{\cal H}}{\partial \vec{p}^{\hspace{0.7mm}2}_n}\delta
\vec{p}_n+\frac{\partial^2{\cal H}}{\partial \vec{p}_n\partial \vec{z}_n}\delta
\vec{z}_n\nonumber\\
-\delta\dot{\vec{p}_n}&=&\frac{\partial^2{\cal H}}{\partial \vec{z}_n\partial
\vec{p}_n}\delta
\vec{p}_n+\frac{\partial^2{\cal H}}{\partial \vec{z}_n^{\hspace{0.7mm}2}}\delta \vec{z}_n\;,
\end{eqnarray}
and using the notation
\begin{eqnarray}
{\cal J}=\left(\begin{array}{cc} 0 & E\\-E & 0 \end{array} \right)\;,
\end{eqnarray}
where $E$ is the identity matrix, we obtain
\begin{eqnarray}\label{b-12}
\left(\begin{array}{c}\delta \dot{\vec{p}}_n\\ \delta \dot{\vec{z}_n}\end{array}\right)={\cal
J}^{-1}\nabla^2{\cal H}\left(\begin{array}{c}\delta {\vec{p}_n}\\ \delta
{\vec{z}_n}\end{array}\right)\;,
\end{eqnarray}
where $\nabla^2{\cal H}$ is the {\it Hessian} of $\cal H$.

Let us consider the following map by integrating the equations (\ref{b-12}) over
one period $T_b$ of the initial solution
\begin{eqnarray}
\left(\begin{array}{c}\delta {\vec{p}_n}(T_b)\\ \delta
{\vec{z}_n}(T_b)\end{array}\right)=U(T_b)\left(\begin{array}{c}\delta {\vec{p}_n}(0)\\ \delta
{\vec{z}_n}(0)\end{array}\right)\;.
\end{eqnarray}
Since the form ${\cal I}$ is symplectic (\ref{sympprod})
, $U(T_b)$ is symplectic too. 
As a result, we find that the eigenvalues of $U(T_b)$ have to fulfill
the condition that if $\nu$ is an  eigenvalue then
$\frac{1}{\nu}$, $\nu^*$
and $\frac{1}{\nu^*}$ are also eigenvalues. Note that
for a marginally stable periodic motion of a Hamiltonian system  
the Floquet eigenvalues $\nu$ are located on the unit circle.
Switching to an unstable state is realized by collisions of eigenvalues
on the unit circle and departing from it. 

All of the obtained relations for eigenvalues (and eigenvectors) can be rewritten 
for the original
$\delta_n (t)$ variables (call the corresponding Floquet eigenvalues
$\mu$). First of all, we expect that most of the eigenvalues
will be located
on a circle of radius (\ref{radius}). 
Further it follows
that if $\mu$ is an eigenvalue then 
$\frac{e^{-\alpha T_b}}{\mu}$, $\mu^*$ and  $\frac{e^{-\alpha T_b}}{\mu^*}$
are eigenvalues too.
So, all scenarios of collisions of the eigenvalues 
are similar to the Hamiltonian case.
For a periodic motion to be stable we need all eigenvalues $\mu$
to have absolute values less or equal to one, i.e. the complex
numbers $\mu$ should reside inside the unit circle in the complex
plane. 
Since the radius $R(\alpha)\le 1$  it needs further finite variation
of the control parameters (after a collision) 
to enforce an eigenvalue to traverse the finite
distance to the unit circle and to exit it. Thus not every collision
will lead to an instability. 
However, the decrease of the dissipation parameter 
$\alpha$ may tune the system closer to the Hamiltonian case,
making its states more sensitive to any occurring collisions.
Note that due to the periodicity of the breather state 
there is always one eigenvalue $\mu=1$, whose eigenvector
is tangent to the breather orbit. Consequently we always find another
eigenvalue $\mu=e^{-\alpha T_b}$ which is located on the positive real
axis inside the unit circle. During all numerical computations
of Floquet eigenvalues and eigenvectors the above properties were
tested and complete agreement was found.


\begin{thebibliography}{10}
\bibitem{comment2}
\bibinfo{journal}{Similar switchings with an increasing 
  number of resistive junctions have been obtained in [Ref.\onlinecite{rtgfvk01}] 
  for a rather different set of dynamic equations. Such
  switchings have not been found in extensive numerical simulations in [Ref.
  \onlinecite{etjjmabtpo00}]. 
  We think that this difference is due to the different step
  size of the dc bias used. We found that one has to use
  sufficiently small step sizes in order to observe these switchings.}
\bibitem{locmodenote}
\bibinfo{journal}{Note that the frequencies of these two localized EEs can be
nicely predicted taking into account the dynamics of two
cells around the resistive breather center. This will be published
elsewhere.}
\end{thebibliography}
\end{document}